%% file: main.tex
\documentclass[aps,prb,longbibliography,twocolumn,superscriptaddress]{revtex4-2}
\usepackage{amsmath,amssymb,bm,graphicx,color,bbold,hyperref,keyval,url,latexsym,dsfont}
\usepackage{xcolor,mathtools}
\usepackage[normalem]{ulem}
\usepackage{CJK}
\usepackage{float}
\usepackage{comment}
\usepackage{tikz}
\usetikzlibrary{shapes.misc}
\newcommand{\chitr}{\chi_{\rm tr}}

\begin{document}
\title{Site Basis Excitation Ansatz for Matrix Product States}
\author{Steven R. White}
\affiliation{Department of Physics and Astronomy, University of California, Irvine, California 92697, USA} \begin{abstract}
We introduce a simple and efficient variation of the tangent-space excitation ansatz used to compute elementary excitation spectra of one-dimensional quantum lattice systems using matrix product states (MPS). A small basis for the excitation tensors is formed based on a single diagonalization analogous to a single site DMRG step but for multiple states. Once overlap and Hamiltonian matrix elements are found, obtaining the excitation for any momentum only requires diagonalization of a tiny matrix, akin to a non-orthogonal band-theory diagonalization. The approach is based on an infinite MPS description of the ground state, and we introduce an extremely simple alternative to variational uniform matrix product states (VUMPS) based on finite system DMRG. For the $S=1$ Heisenberg chain, our method---site basis excitation ansatz (SBEA)---efficiently produces the one-magnon dispersion with high accuracy. We also examine the role of MPS gauge choices, finding that not imposing a gauge condition---leaving the basis nonorthogonal---is crucial for the approach,  whereas imposing a left-orthonormal gauge (as in prior work) severely hampers convergence.  We also show how one can construct Wannier excitations, analogous to the Wannier functions of band theory, where one Wannier excitation, translated to all sites, can reconstruct the single magnon modes exactly for all momenta.

\end{abstract} 
\maketitle 
The spectrum of elementary excitations of a quantum system is of fundamental importance both for theoretical understanding and for direct connection with experiments. Spectral functions are generally more difficult to obtain compared to static properties, and a number of numerical approaches have been utilized in the past, such as small-system exact diagonalization combined with Lanczos/continued-fraction evaluation of dynamical correlators\cite{GaglianoBalseiro1987,GaglianoDagottoMoreoAlcaraz1986}, or cluster perturbation theory, which extends exact diagonalization with perturbation theory\cite{Senechal2000,Senechal2002}. For a one-dimensional system, conventional
density matrix renormalization group (DMRG)\cite{dmrg1,dmrg2} methods allow one to obtain a limited number of excited states to be obtained, but obtaining a full dispersion relation is difficult. Dynamical algorithms, such as dynamical DMRG\cite{Jeckelmann2002}, which works in frequency space, or those that use real-time evolution\cite{Vidal2004, Daley2004, WhiteFeiguin2004,Haegeman2016,Schollwoeck2011,Schmitteckert2004}, offer an alternative\cite{WhiteAffleck08}. There is current great interest in applying such methods to more challenging two-dimensional systems, which may be at the limits of current computational capability.  A dramatic increase
in efficiency arose with the introduction of the matrix product state (MPS) version of the single-mode approximation, called the excitation ansatz (EA)\cite{HaegemanPRB12,VanderstraetenLecture19}. In EA, the excitation is a plane-wave superposition of MPS, where in each MPS the tensor on one site is different from that of the ground state and creates a local excitation near that site. Remarkably, the accuracy of this simple ansatz for simple gapped systems often rivals that of DMRG itself.

On a technical level, this approach is not without difficulties.  The standard method for obtaining a uniform infinite MPS, the variational uniform matrix product states\cite{VUMPS18} (VUMPS) method, can fail to converge depending on the initial matrices, and the original formulation involved a complicated set of diagrams even for near-neighbor interactions\cite{HaegemanPRB12}. Although EA is very efficient compared to previous approaches, each different $k$ requires a separate full diagonalization.  Our site basis excitation ansatz (SBEA) addresses each of these shortcomings, increasing the efficiency of an already-efficient method. Perhaps just as important, our approach is arguably simpler, lowering the threshold for adoption.  

We will test our algorithms using the spin $S=1$ antiferromagnetic Heisenberg chain, with Hamiltonian
\begin{equation}
H = J \sum_j \mathbf{S}_j \cdot \mathbf{S}_{j+1},
\label{eq:heisenberg}
\end{equation}
where $\mathbf{S}_j$ are the spin-1 operators at site $j$ and $J=1$. This model is gapped\cite{Haldane1983a, Haldane1983b}, and the lowest excitations are triplet magnons\cite{Affleck1987, Affleck1989, WhiteHuse93}. This well-understood model serves as a benchmark and development platform, rather than a system where we obtain new insight. We hope that our developments will serve to facilitate studies of more difficult systems, particularly in two dimensions.  Two of the model's features, being an integer spin chain and one with a finite gap and correlation length ($\xi \approx 6$), are important for the algorithm used.  The finite correlation length means that finite system DMRG can reach the bulk in the center of a finite system. The $\ldots AAA \ldots$ single-site infinite MPS form is appropriate for integer spin chains.  (Here and below, we will frequently use this simplified notation for MPS, where here the notation indicates that all of the tensors are identical in a uniform MPS.  It is to be understood that the tensors $A$ are matrices in their connections to their neighbors, but that they also have a third index, which denotes the physical states of a site.) Note that we will make use of the finite (and modest) correlation length to utilize certain algorithmic simplifications---shortcuts, which are not central to SBEA, but which may be very convenient in avoiding some technical complications particularly for the non-expert.

\emph{The uniform infinite ground state MPS.---} The 
first step is to obtain a uniform or infinite MPS (iMPS) 
representation of the ground state. Substantial progress has been made in this area, and two powerful methods are infinite DMRG (iDMRG)\cite{McCulloch08} and the variational uniform MPS algorithm (VUMPS)\cite{VUMPS18}.   Here our main goal in looking for an alternative was simplicity, rather than pure efficiency.  Finite system DMRG\cite{dmrg1,dmrg2} is well understood, efficient, and highly reliable, and the finite-size gaps of the system help speed convergence. If we take finite system DMRG as a black box, can we get uniform infinite MPS?
It turns out that answer is yes, and the method is almost trivially simple. It also does not appear to be costly compared to existing infinite methods. It can be considered a simplification of iDMRG, but without utilizing recursion.

We start by performing finite system
DMRG on a long open chain ($N \gg \xi$, e.g. $N = 150$), where a matrix product operator (MPO) form for the Hamiltonian is convenient, particularly since MPOs can be generated automatically\cite{itensor}. For $N \gg \xi$ the center is representative of the infinite system in its local properties, but the central tensors have random gauges.  The random gauges stem from the fact that inserting on any link $M M^{-1}$, with $M$ an ordinary nonsingular matrix, leaves the wavefunction unchanged.  Left or right orthogonality of each tensor, which DMRG requires away from the orthogonality center, does not resolve this gauge ambiguity, and the $\ldots ABCD \ldots$ form interferes with forming the desired $\ldots AAA \ldots$ infinite uniform MPS. Specifically, we can say the states of the links on each side of each tensor are different, whereas for the $\ldots AAA \ldots$ form they should be the same.  

To make the links the same, we insert a new site in the middle, with a new tensor. (Inserting sites in the middle goes all the way back to the original infinite-system DMRG\cite{dmrg1,dmrg2}.) To accommodate the new  site, we simply reconstruct the Hamiltonian MPO for $N+1$ sites. Picking a central link on the $N$-site system, we put the orthogonality center there, and obtain a $\tilde \lambda$ diagonal matrix of Schmidt coefficients for that link. (For example, to put it between tensors $CD$, we contract these two tensors and then split them apart with a singular value decomposition, obtaining $C' \tilde \lambda D'$.) The half-MPS on either side of the link form orthogonal bases for $\tilde \lambda$.  We add a new site at that link, replacing $\tilde \lambda$ with a new three leg tensor $\tilde D$, which we initialize randomly.  We then perform a Lanczos diagonalization to optimize the energy, to high accuracy, equivalent to a single step of single site finite-size DMRG, but without a good initial state.  Since this is done only once, the extra cost of the poor initialization and high accuracy is negligible.
As in the original infinite system DMRG, the overall state obtained (assuming a single site unit cell is appropriate) is just as accurate for $N+1$ sites as the state it started from was on $N$ sites. And, importantly, $\tilde D$ has identical left and right indices, so it can immediately make an infinite MPS.

The canonical form of an infinite MPS looks like $\ldots \lambda \Gamma\lambda \Gamma \ldots$,
where $\Gamma$ have the site indices and $\lambda$ is a diagonal matrix of Schmidt coefficients. The left half-MPS $\ldots \lambda \Gamma\lambda \Gamma$ is left-orthogonal and
the right half-MPS $\Gamma\lambda \Gamma\lambda\ldots$ is right orthogonal. 
Since $\tilde D$ sits between two orthogonal half-chains, and it has one site, it must play the role of $\lambda \Gamma \lambda$.
We form $\tilde \Gamma = \tilde \lambda^{-1} \tilde D \tilde\lambda^{-1}$.  We now have a $\ldots \tilde \lambda \tilde \Gamma \tilde \lambda \tilde \Gamma \ldots$ form of a uniform infinite MPS.  

We still do not have the optimal gauge defining the canonical form, where each link corresponds to infinite system Schmidt states; but we can perform the standard Orus and Vidal procedure to obtain the canonical form\cite{OrusVidal2008}. At this point we have two tensors, $\lambda$ and $\Gamma$, describing the ground state, as
$\ldots \lambda \Gamma\lambda \Gamma \ldots$. An attractive way of writing this (in terms of shorter formulas) is to write $A = \lambda^{1/2} \Gamma \lambda^{1/2}$.  Then the MPS is written as $\ldots AAA \ldots$.  We call this the symmetric form.  To make an orthogonal basis on the left, we would use $\ldots AAA\lambda^{-1/2}$, and similarly for the right.

Note that the cost for this procedure is almost completely the cost of the first finite-size DMRG, $O(\xi \chi^3)$, where $\chi$ is the bond dimension, and neglecting the MPO and site dimensions.  The multiple sweeps for finite system DMRG do not matter much, as one uses a smaller $\chi$ for earlier sweeps.   For the $S=1$ Heisenberg chain, we obtain essentially exact infinite ground states with this procedure, limited only by double-precision floating point accuracy. Starting from a finite system also clarifies choices of unit cells.  For example, for a half-integer chain, if one conserves $S_z$, each tensor will have half-integer spin values on one side and integer spin values on the other. This makes an $\ldots ABAB\ldots$ form much more natural than an $\ldots AAA \ldots$ form; the $\ldots AAA \ldots$ form would be  \emph{non-injective}, effectively representing a superposition of degenerate states\cite{SpinonHolon18} whose links combined integer and half-integer spins. Clearly, our algorithm can be generalized to larger unit cells, of size $p$.  One would insert $p$ sites, and perform a finite-segment DMRG to optimize $p$ tensors. Here the method would become even more similar to iDMRG, which in its original description used $p=2$\cite{McCulloch08,OsborneMcCulloch2025}. One Orus-Vidal procedure would suffice to find the optimal gauge at the edges of the unit cell, then standard orthogonalization would provide the canonical  $\ldots \lambda_1 \Gamma_1\ldots \lambda_p \Gamma_p \lambda_1 \Gamma_1\ldots$ form. 

In cases with long or infinite correlation lengths, our simple approach may form a good initial state for application of other methods, such as VUMPS.  It would also be very interesting to compare how accurate the simple method is on its own, compared to, say, VUMPS, particularly for gapless systems. For such infinite correlation length systems, the true ground state has a divergent entanglement entropy.  Most methods using infinite uniform MPS control the entanglement through a finite specified bond dimension. In our case, the entanglement would be controlled slightly differently, through the finite entanglement of the $N$-site system utilized. Presumably the two types of states are very similar.

\emph{The site basis excitation ansatz.---}  In the EA\cite{HaegemanPRL11,HaegemanPRB12}, (with unit cell size $p=1$) we consider an infinite MPS where one ground-state $A$ tensor has been replaced by another tensor, $B$, i.e. a $\ldots A A A B A A A\ldots$ form. If the $B$ is at site $j$, we denote this as $|B\rangle_j$. The EA for an excitation with momentum $k$ is 
\begin{equation}
|\Psi_k(B)\rangle = \sum_j e^{i k j}  |B\rangle_j
\end{equation}
where the tensor $B$ is the same no matter which site it is on.  This $B$ is determined by optimizing the excitation for the minimum energy,  separately for each $k$, so that we have $B(k)$,
and we obtain an excitation energy $E(k)$. Although there is only one $B$ to determine for each $k$, the optimization via an effective Hamiltonian is more complicated than in DMRG where each site tensor is independent. Although this is a superposition of MPS, one for each possible position of $B$, it was observed by Osborne and McCulloch \cite{OsborneMcCulloch2025} that this state can also be represented exactly as a single MPS with doubled bond dimension. 

In our site basis excitation ansatz (SBEA) we introduce a minimal approximate basis $\{B_\alpha\}$ for $\alpha=1\ldots N_\alpha$, which can represent all the $B(k)$ simultaneously with appropriate coefficients $c^k_\alpha$.  We write
\begin{equation}
    B(k) = \sum_{\alpha=1}^{N_\alpha} c^k_\alpha B_\alpha
\end{equation}
where $N_\alpha$ is not too large.
The ${B_{\alpha}}$ must span the space of all $B(k)$. To find the optimal ${B_{\alpha}}$, one could imagine computing $B(k)$ for a grid of $k$, vectorizing them, and performing a principal component analysis to find the optimal basis, but this defeats the purpose of avoiding the per-$k$ calculations.

Instead, we make an additional ansatz for the ${B_{\alpha}}$: we obtain them from a single site Lanczos calculation of a single $B$ on one site, similar to that used above in the generation of the infinite ground state, but we use it to obtain the lowest $N_\alpha$ eigenstates of the effective Hamiltonian for the single site, where $N_\alpha$ is set to go up to an energy cutoff.  Why could this be a good ansatz?  The degrees of freedom involved are the states of the site and the Schmidt vectors on either side of the site, extending into the bulk.  The highest probability Schmidt vectors are quite local to the site; lower probability vectors extend farther and farther into the bulk.  If we imagine the Schmidt vectors for a finite bond dimension of the ground state extending a certain distance into the bulk, then as we find more eigenstates we are approximating a complete set of states in the region near the site, but with an energy cutoff.  Then the EA superposes that group of states over all sites, so it is natural to expect a nearly complete basis overall up to the energy cutoff. However, the states involved all have a single $B$, so we expect the ansatz could cover the single excitation space but have the same limitations as EA, e.g. not covering two-particle excitations.  In addition, there will be substantial overlap of these states from nearby sites, so in constructing the full basis over all the sites it is essential to treat the non-orthogonality properly.

Here are the technical details associated with this Lanczos: in order for the states obtained  to properly minimize overall energy,
we should make sure we have the site as an orthogonality center. Specifically,  we define $A_L = A\lambda^{-1/2}$ and $A_R =\lambda^{-1/2} A $, and we will optimize tensors
$\tilde B$ in an MPS of the form $\ldots AA_L \tilde B A_R A\ldots$. For an initial tensor we set  $\tilde B = S^+(\lambda^{1/2} A \lambda^{1/2})$, where the $S^+$ operator acts on the site index of $A$, which has been omitted for simplicity. In our calculations, we conserve total $S_z$ ($U(1)$ symmetry), so that the $S^+$ puts the states in a different $S_z$ sector, automatically orthogonal to the ground state.

We assume the Hamiltonian is expressed in MPO form, and we need edge tensors to express the effective Hamiltonian for $\tilde B$. For convenience we subtract the ground state energy per bond from each bond term, so the ground state has energy 0. For an infinite MPO, the edge tensors are fixed points of a stacked ``AHA" transfer matrix\cite{VUMPS18,VanderstraetenLecture19,OsborneMcCulloch2025}, which can be solved as an eigenvalue problem. Instead, here we use a simplicity shortcut based on the finite correlation length, noting that we only need to generate the two edge tensors once, taking little time.  We choose a particular center site to put the $B$'s. The Hamiltonian terms  connecting sites a distance much greater than $\xi$ (say 100-200) from the center do not affect the results,  and we set them to zero, resulting in a finite MPO which does not have to be in a special infinite-size form and which can be generated automatically even for complicated Hamiltonians. Then a finite size contraction is done up to the center site, giving the left and right edge tensors. For simplicity, consider a MPO segment of (unrealistically) short length three. 
Then the expectation value of this in the ground state is given by the diagram
\begin{center}
\input{inlineHMPO.txt}
\end{center}
\noindent which shows how finite MPO's are to be terminated, and which omits complex conjugation on the top layer for complex wavefunctions.  Following this pattern, the left edge tensor has the form
\begin{center}
\input{inlineLeft.txt}
\end{center}

\noindent but of course, significantly longer, containing all the sites of the finite MPO to the left of the center. A reflected diagram
describes $E_R$.  The effective Hamiltonian diagram for the single site diagonalization has the form
\begin{center}
\input{inlineHs.txt}
\end{center}
where the application of the Hamiltonian $B' = H B$ involves inserting $B$ on the bottom and reading if off on the top.

We then perform Lanczos with this effective Hamiltonian, converging to the set of $N_\alpha$ lowest lying eigenstates.
The second aspect of our ansatz for the $B_\alpha$ is that one should include enough levels so that the maximum eigenenergy we include is at least at the top of the dispersion curve, which is about $2.7J$ in the Heisenberg chain. It is natural that a level near a particular energy would be necessary to describe a momentum state near that energy. Conversely, we argue that states above
the top should not matter. Below, we verify this assertion---but note that it no longer applies if we impose a left-gauge condition.

The number of levels one obtains up to the band maximum depends on the bond dimension used for the ground state.  This is inconvenient---one would like to use the high accuracy easily obtained for the ground state, but not make it drastically increase the effort for the excited states, where one may be satisfied with lower accuracy. However, note that smaller probability singular values describe more extended Schmidt states, but in the EA the $B$'s are already superposed over all sites, so these extended states are not very important.  It is thus convenient to pretruncate the environment of the $B$ tensor before the diagonalization. 
On each side of $B$ we use the $\lambda$ singular values to truncate the bond
to a smaller $\chi$, $\chitr$.  We define $U$ to be an isometry projecting onto the $\chi_{\rm tr}$ dominant Schmidt vectors (with $U^\dagger U$=1), which in practice is a rectangular section of the identity matrix of size $\chi\times \chitr$.  We insert $U U^\dagger$ to the left and to the right of $B$. Instead of $B$ we find eigenvectors  $B' = U^\dagger B U$, with $U$'s attached to the effective Hamiltonian, as shown
\begin{center}
\input{inlineHsU.txt}
\end{center}
Once the $B'$ are found, we go back to the original basis with $B = U B' U^\dagger$.
The eigenvalues now reach the top of the band after a smaller number of states. The hope is that this smaller number of states make nearly as good a basis for the single magnon band as the many more states one obtains up to the cutoff without the pre-truncation. The pre-truncation also dramatically decreases the cost of the Lanczos. The truncation $\chitr$ limits the maximum possible $N_\alpha \le \chitr^2 d$, where $d$ is the site dimension, but typically the energy reaches the band maximum well before this limit. A plot of the energy levels versus $\chitr$ is shown in Fig. 1. As one decreases $\chitr$, the energies rise more rapidly.  The important point, as shown below, is that quite good
results are obtained with a strong pre-truncation, corresponding to the green curve in Fig. 1, and producing a corresponding minimal set of $B_\alpha$, e.g. $\chitr=8$ and $N_\alpha=7$.

 \begin{figure}[t]
\includegraphics[width=0.8\columnwidth]{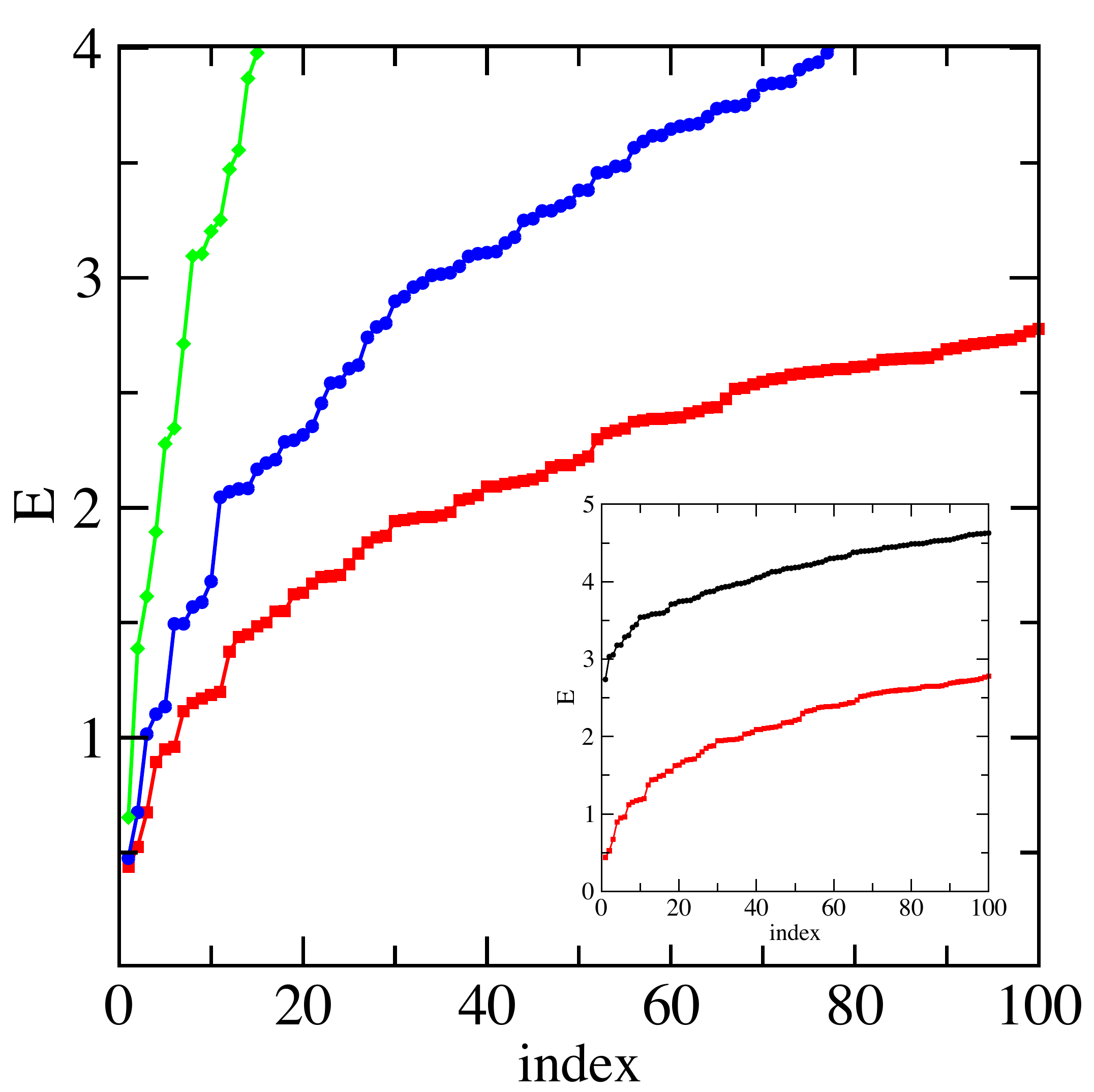}
\caption{Lowest eigenvalues of the effective single-site excitation Hamiltonian $\mathcal{H}$ for the spin-1 chain, obtained via Lanczos. The horizontal axis is the index of the eigenvector (with eigenvalues sorted by increasing energy). Shown are results without pre-truncation (bond dimension $\chi=126$, red circles) and with truncation to $\chi_{\text{tr}}=30$ (blue, with $\lambda^2=10^{-7}$ as a cutoff) and 8 (green, $10^{-4}$).
 The inset shows the untruncated case, where the upper curve
has the left gauge condition imposed, resulting in all the local states being much higher in energy and much less useful.  }
\label{fig:basis_energy}
\end{figure}

\begin{figure}[t]
\includegraphics[width=\columnwidth]{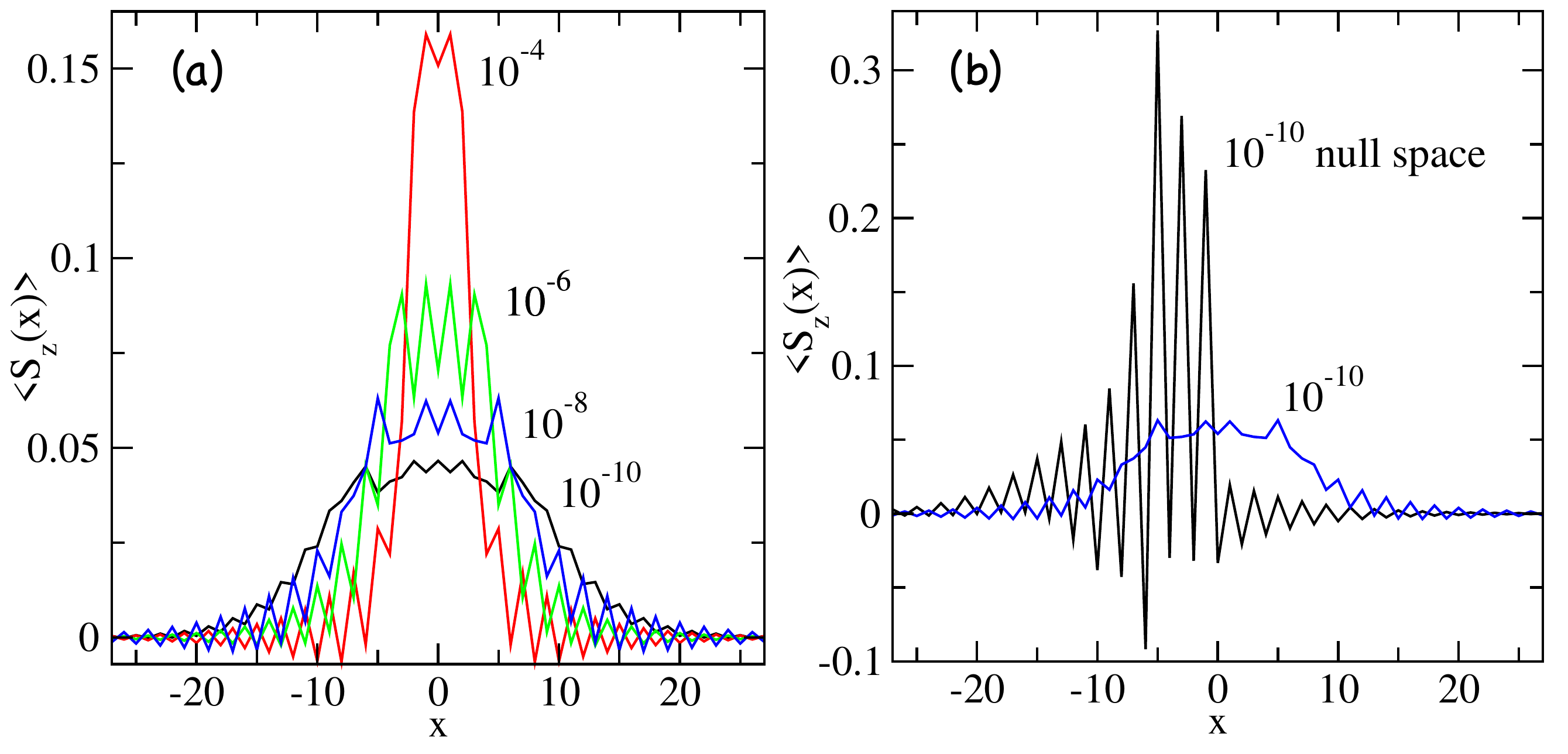}
\caption{Spin profile of single site excitations $\langle S_z(x) \rangle$ versus site $x$. (a) Excitation profiles for the lowest energy single-site state, where a single tensor is different from the ground state.  The numbers beside each curve indicate the cutoff $\lambda^2$ for the pretruncation step; these cutoffs result in $\chi_{\text{tr}}=8, 24, 46$, and $90$ for the curves top to bottom. (b) Comparison of excitation profile for states with and without restriction to the left null space, both with $\lambda^2=10^{-10}$.   }
\label{fig:Szprofile}
\end{figure}

More insight is obtained from observing the spin profiles of the single $B$ states obtained in this process, as shown in Fig. 2.  In part (a), we show the spin profile of the lowest energy single-site state as a function of the pretruncation.  A natural way to obtain a low energy excitation confined to a certain region would be to make a wavepacket say with a Gaussian profile. (The momentum of the packet would be centered at $k=\pi$, the gap minimum point, but this does not affect the profile.) We see that the excitations roughly match such a profile, but the pretruncation decreases the width of the packet. It is natural to expect a superposition of such packets at all  sites to able to represent a pure $k$ state near $k=\pi$. We expect that this superposition would work well even if the packet is fairly narrow. The larger widths would give more accurate dispersions but would also make the basis of such states more linearly dependent.

Normally a left gauge condition is imposed with EA\cite{HaegemanPRB12}, which makes states $|B\rangle_j$ and $|B'\rangle_{j'}$ orthogonal if $j \ne j'$. The left gauge condition
involves projecting into the null spaces associated with each site. The gauge choice can be omitted if one explicitly takes into account the non-orthogonality properly.
It turns out that \emph{not} imposing the left gauge condition is important for the truncation to small $N_\alpha$. As shown in the inset of Fig. 1, with the left gauge condition, none of the eigenvalues are low in energy.  Since the span of the basis combining all sites is not affected by the gauge condition, this means that strong connections between the states with $B$'s on different sites are necessary to obtain the lower part of the spectrum. In Fig. 2(b), we show the spin profile of a single site state obtained with the left gauge (null space) condition, comparing with an equivalent profile from (a). The profile looks much less like a minimum energy wavepacket, and it shows strong left-right asymmetry.  These unphysical features make the left-gauge states much less suitable for truncation to a small basis.
In practice, the SBEA algorithm was not practical with the left gauge condition; $N_\alpha$ would be too large for reasonable accuracy (note that not truncating would correspond to $N_\alpha \sim \chi^2$.)

The number of degrees of freedom in each $B$ is smaller with the left-gauge imposed, $\chi^2$ versus $\chi^2 d$, where $d$ is the site dimension, and not pre-truncating.  Without any truncation, and without the left gauge, the basis would be overcomplete. However, typically our truncations keep us far from this limit.

\newcommand{\braj}{\prescript{}{j\negthinspace}{\langle}}
Having settled on a set of $B_{\alpha}$, which are not orthogonal on different sites $j$, we next evaluate the overlap kernel
$O_{\alpha\alpha'}(j'-j) = \braj B_\alpha | B_{\alpha'}\rangle_{j'}$ and the Hamiltonian kernel $ H_{\alpha\alpha'}(j-j') =  \braj B_\alpha |H| B_{\alpha'}\rangle_{j'}$.
Diagrammatically these quantities are nothing more than these two
sandwiches, illustrated for $j'-j=3$. 

\begin{center}
\input{inlineO.txt}  
\end{center}

\begin{center}
\input{inlineH.txt}  
\end{center}

Because the ground state is gapped, connected correlations decay as
$e^{-|j-j'|/\xi}$ and both kernels are therefore exponentially banded.
In practice we take $-L_c\le j-j'\le L_c$ with $L_c\simeq 150$; beyond that threshold every entry is
below about $10^{-12}$ and can be safely discarded. 
The calculation of the $H$ array of size $N_\alpha^2 L_c$ is the dominant effort in SBEA.  If the Hamiltonian MPO is 
in infinite uniform $\ldots HHH\ldots$ form, then the
contractions are particularly simple, since the edge tensors don't depend on the site.  In that case for each  $B_\alpha$ we start from the left, and progressive contract towards the right.  For each site, there is a final contraction of all the
$B_{\alpha'}$. Then the calculation time is
$O[(N_\alpha^2 + N_\alpha K) L_c \chi^3]$, where $K$ is the MPO bond dimension. The calculation time for the different curves in Fig. 3 varies primarily as $N_\alpha^2$.
This calculation time
is identical to that of a finite-system DMRG calculation on a system of length $L_c$, except the parenthetical factor would be replaced by an effective number of sweeps times the number of Lanczos steps at each step, i.e. still similar to a single DMRG ground state calculation. 

If the Hamiltonian MPO is not in infinite form, then the edge
tensors are different depending on the ending sites of the interval between $j$ and $j'$. In this case,
we make an extra large
region of size $L_c'$, with $L_c' \sim  3 L_c$, and embed the $L_c$ region in the center.  Then edge tensors are calculated by successive contraction from the outer edges to store them for all the sites in the central region. (This is essentially the same as keeping track of edge tensors in a finite-system DMRG calculation.)

\emph{Nonorthogonal band theory.---}Given the overlap and Hamiltonian kernels, we can now solve for the low-lying modes for each momentum $k$, which resembles standard band theory for a tight-binding model with multiple levels per site, but with the orbitals on different sites nonorthogonal.
Fourier transforming the real‑space kernels yields
\begin{equation}
\tilde O_{\alpha\alpha'}(k)=\sum_{d}e^{ikd}O_{\alpha\alpha'}(d)
\end{equation}
and
\begin{equation}
\tilde H_{\alpha\alpha'}(k)=\sum_{d}e^{ikd}H_{\alpha\alpha'}(d).
\end{equation}

For each momentum $k$ we solve the $N_{\alpha}\times N_{\alpha}$ generalized
eigenproblem
\begin{equation}
\tilde H(k)c = E(k)\tilde O(k)c.\label{eq:genEV}
\end{equation}
for eigenvector $c$ and eigenvalue $E(k)$. For large $N_\alpha$, $\tilde O$ could become singular; in such cases we solve the problem in a reduced space with eigenvectors of $\tilde O$ removed whose eigenvalue is below a threshold ($10^{-8}$). However, large $N_\alpha$ is primarily for testing; we don't encounter singular $\tilde O$ for small $N_\alpha$. On a typical Mac laptop, calculating the spectra for 500 different $k$ values
takes a few seconds with $N_\alpha=7$; calculation of $O$ and $H$ beforehand takes less than a minute.

\begin{figure}[t]
\centering
\includegraphics[width=0.8\columnwidth]{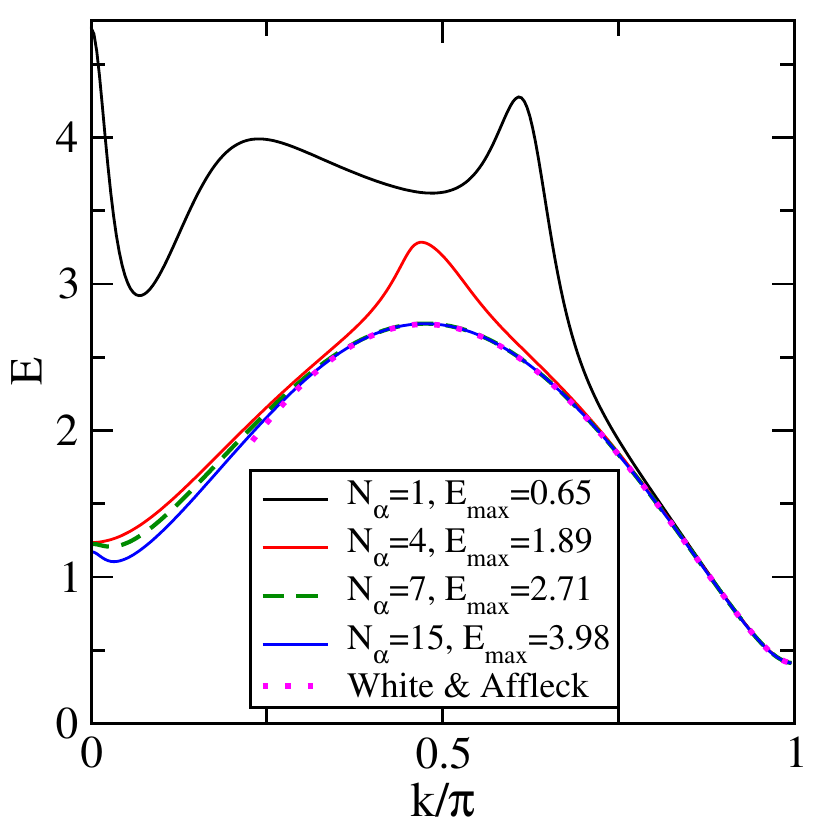}
\caption{One‑magnon dispersion of the spin‑1 Heisenberg chain, as a function of the maximum single-site energy eigenvalue included in
the site-basis.  In all cases, $\chitr=8$; for each curve, $N_\alpha$ of the $B_\alpha$ were used in the basis, with $E_{\rm max}$ indicating the maximum energy.  We see excellent convergence in the spectra for energies
below $E_{\rm max}$, which agree with the tDMRG results of White and Affleck.\cite{WhiteAffleck08}
The one magnon line is inside the two-magnon continuum
for $k$ smaller than about 0.72; in that region, the EA is less appropriate and convergence in $N_\alpha$ is slower.}
\label{fig:dispersion}
\end{figure}

\emph{Results for the $S=1$ Haldane chain.---}Figure \ref{fig:dispersion} shows results for the dispersion $E(k)$ obtained from
Eq.~~\eqref{eq:genEV} (solid line) together with a fit to  results from time‑dependent DMRG\cite{WhiteAffleck08}. 
The number of $B_\alpha$ used affects the results; as expected, the dispersion is inaccurate for energies above the
highest energy $B_\alpha$. Even a single $B_\alpha$ is excellent for the bottom of the band. The dispersion gets progressively better as the last energy included $E_{\rm max}$ moves higher. Excellent results are found for $N_\alpha=7$, which covers the whole dispersion energy curve. 
Note that the dispersion is not defined once the single magnon line enters the two-magnon continuum near $k \approx 0.23\pi\approx 0.72$\cite{WhiteAffleck08}, so the fit is not given for $k < 0.72$. The lowest energy state at $k=0$ is formed from two magnons at $k=\pi$, with an energy of $0.82$. The EA is not adequate to treat these states.  A better approach would be the two-particle EA\cite{Vanderstraeten2015}; alternatively, to track broadened single-magnon excitations one could use the variance reduction ansatz of Osborne and McCulloch\cite{OsborneMcCulloch2025}.
At $k=\pi$ we find for $N_\alpha=7$ a Haldane gap of $\Delta/J=0.4107$, versus the essentially exact result
$\Delta/J=0.410479\ldots$. This small error is comparable to the pretruncation cutoff of $10^{-4}$ used to get $N_{tr}=8$.  We see that SBEA is remarkably efficient overall for obtaining fairly high accuracy spectra for the Haldane chain.

\emph{Excitation space and Wannier Functions.---}
An important mathematical aspect of the excitation ansatz is that the set of states generated by inserting an arbitrary tensor $B_j$ at one site $j$, with all other sites carrying the ground state tensor $A$, combined over all $j$,  forms a complex vector space. The tangent space of an MPS is an important concept underlying time evolution algorithms\cite{Haegeman2016}, but the excitation space is distinct. For example, in symmetric systems the excitation space encompasses states with different quantum numbers, whereas the tangent space, expressing small changes to the ground state, has the same quantum numbers as the ground state.  The elements of this vector space have the form
\begin{equation}
|\{B_j\}\rangle = \sum_{j} |B_j\rangle_j .
\end{equation}
The space of all such single-site excitations, with arbitrary choices of $B_j$ at each site, is closed under addition and scalar multiplication and thus constitutes a vector space. Note also that evaluating linear combinations is very simple, e.g. 
\begin{equation}
b|\{B_j\}\rangle +c|\{C_j\}\rangle = |\{bB_j+cC_j\}\rangle
\end{equation} 
where $b$ and $c$ are scalars. The states from the standard EA live in this space. 
We refer to this space as the \textit{single-site excitation space} associated with a given ground state MPS. 

The vector space structure allows for the straightforward construction of analogs of Wannier functions, paralleling familiar procedures in electronic band theory\cite{MarzariVanderbilt1997,Marzari2012}. This construction has some similarities to previous constructions of localized wavepackets\cite{vandamme2021}.
The Wannier excitations are states of the system that have the following properties:  there is one associated with every site; they are orthonormal; they span all single magnon excitations; and they are local.
We construct a Wannier excitation as a linear combination of momentum eigenstates spanning the magnon band.
To help control the arbitrary phases always involved in the construction of Wannier functions, we form a projection operator out of the momentum resolved single magnon excitations $|k\rangle$:
\begin{equation}
P =  \sum_{k} |k\rangle\langle k| \, .
\end{equation}
We then apply $P$ to a set of identical (except for translation) localized states $|L_j\rangle$. (The arbitrary phases then correspond to arbitrary choices of the $|L_j\rangle$.) To quantize the $k$ levels we put the system on a periodic ring with a circumference substantially larger than the correlation length (e.g. $N = 200$). (The periodic construction is only done after obtaining the $\tilde O$ and $\tilde H$ matrices for an infinite system, so no periodic DMRG is needed.) We
obtain
\begin{equation}
|\tilde W_j\rangle = P |L_j\rangle
\end{equation}
and then symmetrically orthonormalize
\begin{equation}
|W_j\rangle = |\tilde W_{j'}\rangle [S^{-1/2}]_{j'j}
\end{equation}
where $S_{jj'} = \langle \tilde W_j | \tilde W_{j'}\rangle$. 

The $|W_j\rangle$ are then a superposition of single-$B_{j'}$ states.  Assuming the Wannier construction is successful, there should be non-negligible coefficients for $j'$ only within several correlations lengths of $j$. As in the case of the momentum states, this superposition can also be combined into a single MPS with doubled bond dimension. We emphasize that these Wannier excitations are orthonormal, unlike the states $|B_j\rangle_j$, but they are more complicated, involving a superposition of $|B_j\rangle_j$.

The choice of localized states $|L_j\rangle$ is arbitrary, except that they should not be orthogonal (in aggregate) to any of the excitations $|k\rangle$. The choice could be optimized to improve locality\cite{MarzariVanderbilt1997}. A simple choice would be the lowest energy single-site-tensor state, $|L_j\rangle=|B_{\alpha=1}\rangle_j$, but we found that $S$ was singular for this choice, presumably because this single state has a definite parity under reflection. Instead, we tried a simple mixture designed to eliminate definite parity, $B = B_{\alpha=1} + 0.1 \sum_{\alpha \ne 1} B_\alpha$. For this choice, $S$ is nonsingular and the Wannier construction is satisfactory.

The construction is then successful if the Wannier excitations are localized---both in the range with non-negligible $|B_j\rangle_j$ and in the properties, such as the spin profile.  Fortunately, these two aspects are tied together by the locality of the properties of the $|B_j\rangle_j$. The locality is also tied to $S$ not being singular or nearly singular. Although the EA is not well-justified in the two-magnon regime $k < 0.72$, any treatment of this regime that was sharply different from $k > 0.72$ would likely lead to non-localized Wannier excitations.  Therefore, we simply used the states obtained for all $k$, ignoring questions of whether the states were properly single-magnon.
\begin{figure}[t]
\centering
\includegraphics[width=0.7\columnwidth]{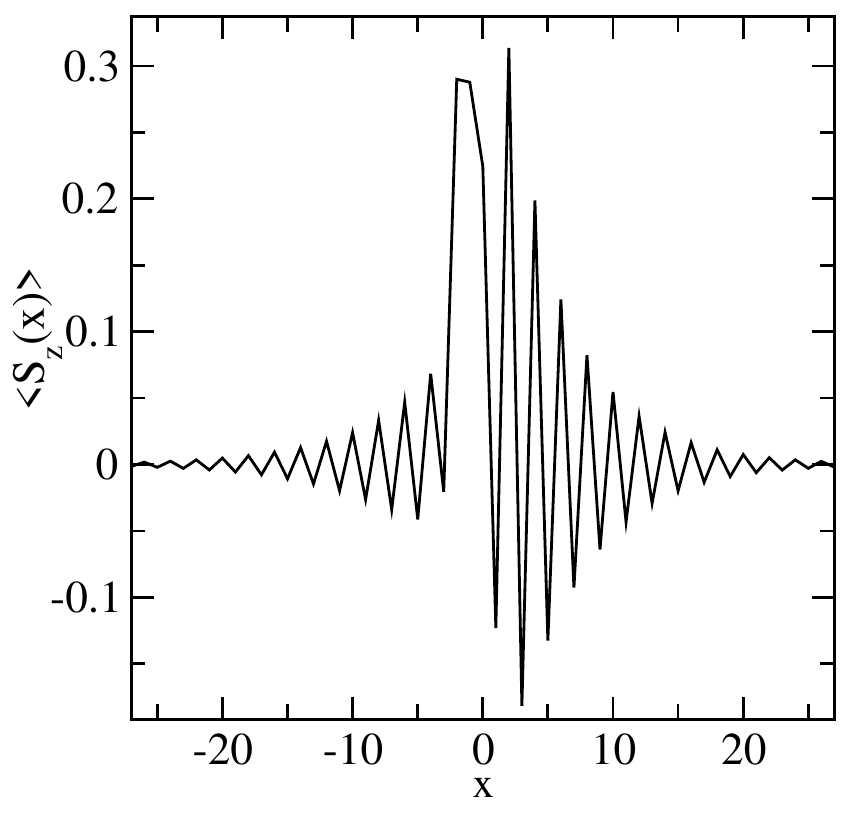}
\caption{Spin profile of a Wannier excitation for the Heisenberg chain, using pretruncation $\chi_{\text{tr}}=8$, and $N_\alpha=7$.  The profile shows that the excitation is well-localized.  Its lack of reflection symmetry is tied to the choice of local states $|L_j\rangle$.}
\label{fig:Wannier}
\end{figure}

In Fig. 4, we show the spin profile for the Wannier excitation. The key point is that it is indeed well-localized, over a range of several times the correlation length.  Without this localization, the Wannier excitations would not be useful.
Note that diagonalizing the Hamiltonian matrix $H_{jj'} = \langle W_j | H | W_j'\rangle$ reproduces exactly the approximate single magnon dispersion obtained with SBEA (not shown). These Wannier excitations could form a starting point for a renormalized multi-magnon effective Hamiltonian, paralleling the use of Wannier functions in building Hubbard-like models. If two particles are far apart, it is clear how to form the double excitation state, but an ansatz would be needed to describe a state when two magnons are nearby.

\emph{Conclusions.---}The MPS excitation approximation, based on infinite MPS  ground states, has remarkable accuracy for such a simple ansatz, when applied to gapped chains such as the $S=1$ Heisenberg chain.  We have revisited this approach, introducing a set of innovations that can make this method more convenient, simpler, and more computationally efficient. First, we point out that infinite MPS can be obtained from finite MPS by a simple site-insertion Lanczos calculation.  In cases where other methods, such as VUMPS, have difficulty with convergence or initialization, this could be particularly useful. We have shown that the EA can be made more computationally efficient by making a site basis derived from another short
Lanczos calculation for a single site, the SBEA method. After constructing Hamiltonian overlaps, the basis then makes the solution for a particular $k$ a trivial generalization of non-orthogonal band theory.  A key ingredient to this is avoiding the standard left-gauge choice; at the cost of non-orthogonality, we can sharply truncate the size of our basis.   We have also shown how one can define Wannier excitations, analogous to Wannier functions, which are localized and reproduce the single magnon band precisely.  These may prove useful in creating renormalized Hamiltonians to study multi-magnon excitations, encapsulating the single magnon behavior.
Our framework should extend straightforwardly to larger unit
cells, where it will be much more broadly applicable.

\emph{Acknowledgments---.} Supported by the U.S. NSF under Grant
DMR‑2412638.  I thank Daniel White, Miles Stoudenmire, Matt Fishman, Lukas Devos, Thomas Scaffidi, Ian McCulloch, and Jesse Osborne
for very helpful discussions.  The calculations were performed using the ITensor library\cite{itensor}.

\bibliography{ref}
\end{document}

%% file: inlineHMPO.txt
\begin{tikzpicture}[
        scale=0.5, transform shape,          
        rect/.style={font=\Large},
        tensor/.style={draw,circle,minimum size=20pt,inner sep=0pt,font=\Large},
        phys/.style  ={draw,circle,minimum size=20pt,inner sep=0pt,font=\Large},
        >=latex]

\def\yTop{ 1.2}
\def\yMid{ 0  }
\def\yBot{-1.2}

\node[phys] (l) at (0,0)  {$\lambda$};
\node[phys] (r) at (8,0) {$\lambda$};

\foreach \x in {2,4,6}{
}

\node[tensor] (A1)  at ( 2,\yTop) {$A$};
\node[tensor] (A2)  at ( 4,\yTop) {$A$};
\node[tensor] (A3)  at ( 6,\yTop) {$A$};

\node[tensor] (H1) at ( 2,\yMid) {$H$};
\node[tensor] (H2) at ( 4,\yMid) {$H$};
\node[tensor] (H3) at ( 6,\yMid) {$H$};

\node[tensor] (A4)  at ( 2,\yBot) {$A$};
\node[tensor] (A5)  at ( 4,\yBot) {$A$};
\node[tensor] (A6)  at ( 6,\yBot) {$A$};

\draw (A1) -- (A2) -- (A3);
\draw (H1) -- (H2) -- (H3);
\draw (A4) -- (A5) -- (A6);

\draw (A1)--(H1);
\draw (A2)--(H2);
\draw (A3)--(H3);
\draw (A4)--(H1);
\draw (A5)--(H2);
\draw (A6)--(H3);

\draw (l) to[out=60,in=180] (A1);
\draw (l) to[out=-60,in=180] (A4);
\draw (A3) to[out=0,in=120]  (r);
\draw (A6) to[out=0,in=-120] (r);
\end{tikzpicture}

%% file: inlineLeft.txt
\begin{tikzpicture}[
scale=0.5, transform shape,          
rect/.style={font=\Large},
tensor/.style={draw,circle,minimum size=20pt,inner sep=0pt,font=\Large},
phys/.style  ={draw,circle,minimum size=20pt,inner sep=0pt,font=\Large},
>=latex]

\def\yTop{ 1.2}
\def\yMid{ 0  }
\def\yBot{-1.2}

\draw (-3.4, 1.6) rectangle (-2.0,-1.6);   
\node[font=\Large] at (-2.7,0) {$E_L$};

\draw (-2.0,\yTop) -- (-1.3,\yTop);
\draw (-2.0,0.0) -- (-1.3,0.0);
\draw (-2.0,\yBot) -- (-1.3,\yBot);

\node[font=\Large] at (-0.5,0) {$=$};

\node[tensor] (l) at (0.5,0) {$\lambda$};

\node[tensor] (A1)  at ( 2,\yTop) {$A$};
\node[tensor] (A2)  at ( 4,\yTop) {$A$};
\node[tensor] (A3)  at ( 6,\yTop) {$A$};

\node[tensor] (H1) at ( 2,\yMid) {$H$};
\node[tensor] (H2) at ( 4,\yMid) {$H$};
\node[tensor] (H3) at ( 6,\yMid) {$H$};

\node[tensor] (A4)  at ( 2,\yBot) {$A$};
\node[tensor] (A5)  at ( 4,\yBot) {$A$};
\node[tensor] (A6)  at ( 6,\yBot) {$A$};

\draw (A1) -- (A2) -- (A3);
\draw (H1) -- (H2) -- (H3);
\draw (A4) -- (A5) -- (A6);

\draw (A1)--(H1);
\draw (A2)--(H2);
\draw (A3)--(H3);
\draw (A4)--(H1);
\draw (A5)--(H2);
\draw (A6)--(H3);

\draw (l) to[out=60,in=180] (A1);
\draw (l) to[out=-60,in=180] (A4);
\draw (A3) -- (7.0,\yTop);
\draw (H3) -- (7.0,0.0);
\draw (A6) -- (7.0,\yBot);
\end{tikzpicture}

%% file: inlineHs.txt
\begin{tikzpicture}[
        scale=0.5, transform shape,          
        rect/.style={font=\Large},
        tensor/.style={draw,circle,minimum size=20pt,inner sep=0pt,font=\Large},
        btensor/.style={circle,minimum size=20pt,inner sep=0pt,font=\Large},
        >=latex]

\def\yTop{ 1.2}
\def\yMid{ 0  }
\def\yBot{-1.2}

\draw (-0.7, 1.8) rectangle (0.7,-1.8);   
\node[rect] at (0,0) {$E_L$};

\draw ( 3.3, 1.8) rectangle (4.7,-1.8);  
\node[rect] at (4,0) {$E_R$};

\node[btensor] (A1)  at ( 2,\yTop) {};
\node[tensor] (H1) at ( 2,\yMid) {$H$};
\node[btensor] (A2)  at ( 2,\yBot) {};

\draw (0.7,\yTop) -- (A1) -- (3.3,\yTop);
\draw (0.7,\yMid) -- (H1) -- (3.3,\yMid);
\draw (0.7,\yBot) -- (A2) -- (3.3,\yBot);

\draw (A1) -- (H1) -- (A2);
\end{tikzpicture}

%% file: inlineHsU.txt
\begin{tikzpicture}[
        scale=0.5, transform shape,          
        rect/.style={font=\Large},
        tensor/.style={draw,circle,minimum size=20pt,inner sep=0pt,font=\Large},
        btensor/.style={circle,minimum size=20pt,inner sep=0pt,font=\Large},
        >=latex]

\def\yTop{ 1.2}
\def\yMid{ 0  }
\def\yBot{-1.2}
\def\xa{1.5}
\def\xb{2.5}
\def\xc{3.5}

\draw (-0.7, 1.8) rectangle (0.7,-1.8);   
\node[rect] at (0,0) {$E_L$};

\draw ( 4.3, 1.8) rectangle (5.7,-1.8);  
\node[rect] at (5,0) {$E_R$};

\node[btensor] (A1)  at ( \xb,\yTop) {};
\node[tensor] (ur1) at ( \xc,\yTop) {$U^\dagger$};
\node[tensor] (ur2) at ( \xc,\yBot) {$U^\dagger$};
\node[tensor] (ul1) at ( \xa,\yTop) {$U$};
\node[tensor] (ul2) at ( \xa,\yBot) {$U$};

\node[tensor] (H1) at ( \xb,\yMid) {$H$};
\node[btensor] (A2)  at ( \xb,\yBot) {};

\draw (0.7,\yTop) -- (ul1) -- (A1) -- (ur1) -- (4.3,\yTop);
\draw (0.7,\yMid) -- (H1) -- (4.3,\yMid);
\draw (0.7,\yBot) -- (ul2) -- (A2) -- (ur2) -- (4.3,\yBot);

\draw (A1) -- (H1) -- (A2);
\end{tikzpicture}

%% file: inlineO.txt
\begin{tikzpicture}[
        scale=0.5, transform shape,      
        tensor/.style={draw,circle,minimum size=20pt,inner sep=0pt,font=\Large},
        phys/.style  ={draw,circle,minimum size=20pt,inner sep=0pt,font=\Large},
        >=latex]

\node[phys] (l) at (0,0)  {$\lambda$};
\node[phys] (r) at (10,0) {$\lambda$};

\node[tensor] (t1) at (2, 1.2) {$A$};
\node[tensor] (t2) at (4, 1.2) {$A$};
\node[tensor] (t3) at (6, 1.2) {$A$};
\node[tensor] (t4) at (8, 1.2) {$B_{\alpha'}$};

\node[tensor] (b1) at (2,-1.2) {$B_{\alpha}$};
\node[tensor] (b2) at (4,-1.2) {$A$};
\node[tensor] (b3) at (6,-1.2) {$A$};
\node[tensor] (b4) at (8,-1.2) {$A$};

\draw (t1)--(t2)--(t3)--(t4);
\draw (b1)--(b2)--(b3)--(b4);

\draw (t1)--(b1);
\draw (t2)--(b2);
\draw (t3)--(b3);
\draw (t4)--(b4);

\draw (l) to[out=60,in=180] (t1);
\draw (l) to[out=-60,in=180] (b1);
\draw (t4) to[out=0,in=120]  (r);
\draw (b4) to[out=0,in=-120] (r);
\end{tikzpicture}

%% file: inlineH.txt
\begin{tikzpicture}[
        scale=0.5, transform shape,          
        rect/.style={font=\Large},
        tensor/.style={draw,circle,minimum size=20pt,inner sep=0pt,font=\Large},
        >=latex]

\def\yTop{ 1.2}
\def\yMid{ 0  }
\def\yBot{-1.2}

\draw (-0.7, 1.8) rectangle (0.7,-1.8);   
\node[rect] at (0,0) {$E_L$};

\draw ( 9.3, 1.8) rectangle (10.7,-1.8);  
\node[rect] at (10,0) {$E_R$};

\foreach \x in {2,4,6,8}{
}

\node[tensor] (A1)  at ( 2,\yTop) {$A$};
\node[tensor] (A2)  at ( 4,\yTop) {$A$};
\node[tensor] (A3)  at ( 6,\yTop) {$A$};
\node[tensor] (Bap) at ( 8,\yTop) {$B_{\alpha'}$};

\node[tensor] (H1) at ( 2,\yMid) {$H$};
\node[tensor] (H2) at ( 4,\yMid) {$H$};
\node[tensor] (H3) at ( 6,\yMid) {$H$};
\node[tensor] (H4) at ( 8,\yMid) {$H$};

\node[tensor] (Bal) at ( 2,\yBot) {$B_{\alpha}$};
\node[tensor] (A4)  at ( 4,\yBot) {$A$};
\node[tensor] (A5)  at ( 6,\yBot) {$A$};
\node[tensor] (A6)  at ( 8,\yBot) {$A$};

\draw (0.7,\yTop) -- (A1) -- (A2) -- (A3) -- (Bap) -- (9.3,\yTop);
\draw (0.7,\yMid) -- (H1) -- (H2) -- (H3) -- (H4) -- (9.3,\yMid);
\draw (0.7,\yBot) -- (Bal) -- (A4) -- (A5) -- (A6) -- (9.3,\yBot);

\draw (A1) -- (H1) -- (Bal);
\draw (A2) -- (H2) -- (A4);
\draw (A3) -- (H3) -- (A5);
\draw (Bap)-- (H4) -- (A6);
\end{tikzpicture}